\documentclass[prl,twocolumn,showpacs,twocolumngrid,superbib]{revtex4}
\newcommand{\commentoutB}[1]{}
\newcommand{\commentoutA}[1]{#1}
\usepackage{graphicx}
\usepackage{amsfonts}
\usepackage{amsmath}
\usepackage{bm}
\usepackage{alltt}
\usepackage{fancyhdr}

\begin{document}

\title{A New View on Geometry Optimization: the Quasi-\\
       Independent Curvilinear Coordinate Approximation\footnotemark[1]}

\author{K\'aroly N\'emeth\footnotemark[2]}
\author{Matt Challacombe}

\affiliation{Theoretical Division,\\ Los Alamos National Laboratory, \\ Los Alamos, NM 87545, USA}

\date{\today}

\begin{abstract}
This article presents a new and efficient alternative to well established
algorithms for molecular geometry optimization.   The new approach 
exploits the approximate decoupling of molecular energetics in a curvilinear 
internal coordinate system,  allowing separation  of the 3$N$-dimensional
optimization problem into an ${\cal{O}}(N)$ set of quasi-independent  
one-dimensional problems.  Each  uncoupled optimization is developed by 
a weighted least squares fit of energy gradients in the internal coordinate 
system followed by extrapolation.  In construction of the weights, only an 
implicit dependence on topologically connected internal coordinates is 
present.   This new approach is competitive with the best internal coordinate
geometry optimization algorithms in the literature and works well for large 
biological problems with complicated hydrogen bond networks and ligand binding motifs.\\[.1cm]
\end{abstract}


\maketitle

\footnotetext[1]{LA-UR-04-1097}
\footnotetext[2]{\tt KNemeth@LANL.Gov}

\section{Introduction}

Recently, $N$-scaling electronic structure algorithms ($N$ is the system size)  are beginning to 
realize their promise, delivering energies and forces on nuclei for large systems at 
both a numerical accuracy and theoretical level suitable to begin addressing many complex 
problems.  For example, it is now possible to routinely apply hybrid Hartree-Fock/Density 
Functional theory to fragments of biological molecules involving 
several hundred atoms using basis sets of triple-zeta plus polarization quality.  
Despite favorable scaling properties, however, ${\cal O}(N)$ methods typically carry a larger 
cost prefactor than their conventional counterparts.  Thus, 
in addition to parallelism, developing efficient algorithms to drive the core engines of 
linear scaling electronic structure theory is paramount.  Of these, perhaps the most useful 
driver of electronic structure theory is the molecular geometry optimizer, which follows 
the energy gradient downhill to a local minimum.   

We present a new concept for the optimization of molecular geometries based on
the approximate separability of molecular motions in curvilinear internal coordinate 
systems, and the recent availability of fast transformations to and from these coordinate systems 
for large molecules \cite{paizs_coordtrf1,nemeth_coordtrf1,paizs_coordtrf2,nemeth_coordtrf2}.  
Internal coordinates represent the internal motions of a molecule, involving  bond stretches, angle 
bendings and dihedral torsions.  These motions correspond to different  energy and length 
scales, leading to strong diagonal dominance in the matrix of second order energy derivatives 
(the Hessian) as well as in higher order anharmonic energy tensors.  Off diagonal elements of the 
internal coordinate Hessian are typically an order of magnitude smaller than the diagonal ones
\cite{pulay_69,fogarasi_diaghess,Pulay_natural_internals,pulay_review,pulay_dynamics}.
This is the basis for an approximate decoupling of molecular motions in terms of internal coordinates.

\commentoutA{

\begin{figure}[h]
\resizebox*{3.5in}{!}{\includegraphics{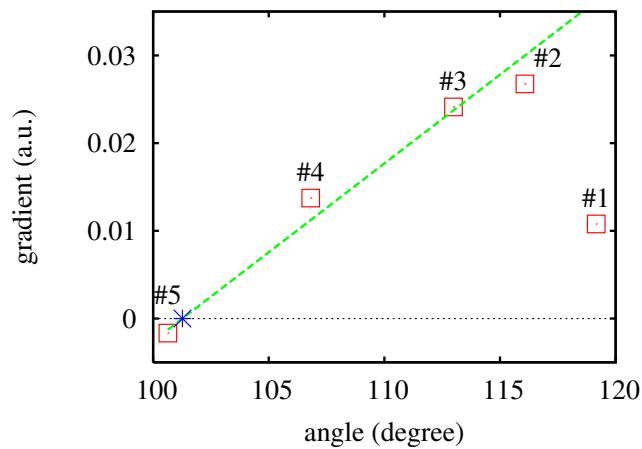}}
\caption{
Progression of gradients on a valence angle bending coordinate of
ammonia. Energies and forces have been obtained by the PBE 
density functional model in STO-3G basis set.
The optimization was started from near planar geometry, i.e.
from the vicinity of a transition state, and converged to a local 
minimum of the potential energy surface. The numbers in the picture
indicate the sequence of optimization steps. The dashed line represents
a linear fit, the star the predicted location of the minimum.}
\label{NH3outp6}
\end{figure}

}

An observation central to the present contribution is that internal coordinate gradients show trends 
during optimization that can be captured by curve fitting and used to predict minima by 
one-dimensional extrapolation.  This is illustrated in Fig.~\ref{NH3outp6}, showing a typical progression of 
internal coordinate gradients.  

In the following, we review conventional methods of geometry optimization (Section II), 
and then  develop our new idea conceptually (Section III).  A practical implementation of this 
concept is then outlined (Section IV) in the context of linear scaling quantum 
chemistry.  Next, this implementation is tested against Baker's small molecule test suite and applied to 
a reaction model of Protein Kinase A (Section V).  

\subsection{Conventional Methods}\label{conventional}

Over the last three decades, internal coordinate geometry optimizers have become the standard 
tools for finding local minima when energies and forces are calculated by electronic structure 
theory. Internal coordinates describe the internal motions of molecules 
with chemical degrees of freedom such as bond-stretches, valence angle bendings and the torsions 
of dihedral angles. These curvilinear coordinates are constructed from non-linear functions of the 
Cartesian coordinates of atomic nuclei \cite{wilson}.

The reduced vibrational coupling achieved with internal coordinates provides an advantage 
in molecular geometry optimization \cite{pulay_review} and molecular dynamics \cite{pulay_dynamics}.
Specifically, a full Newton step in Cartesian coordinates will typically generate a larger, 
anharmonic coupling relative to a full Newton step carried out in internal coordinates.
An excellent discussion of this issue can be found in the Introduction of Ref.~\onlinecite{Pulay_natural_internals}.

Today, the most aggressive internal coordinate algorithms have been highly optimized for 
small molecule quantum chemistry.  These methods typically generate a Cartesian Hessian,
either exactly or approximately. This Cartesian Hessian is then transformed into an internal 
coordinate representation, and an internal coordinate Newton step is taken. The internal 
coordinate step is then transformed back to a Cartesian displacement and used to increment the 
molecular geometry.

There are many variations on this approach.  Cartesian Hessians are usually calculated by updating 
algorithms in the framework of the variable metric or quasi-Newton approach such as the  
Broyden-Fletcher-Goldfarb-Shanno (BFGS) method \cite{RFletcher}.  With increasing system size, storing,
transforming and potentially inverting a full Hessian matrix scales as ${\cal O}(N^3)$, and it becomes
favorable to consider diagonal or semi-diagonal approximations to the internal coordinate Hessian. 
This severely truncated Newton or  ``force relaxation'' method typically obtains static diagonal values 
{\em a priori} from spectroscopic data or empirical force-fields 
\cite{pulay_69,fogarasi_diaghess,Pulay_natural_internals,pulay_review,sellers,van_alsenoy_98,lindh}.

It is worth noting that the landmark RHF/4-21G optimization of Crambin by Van Alsenoy \cite{van_alsenoy_98}
was carried out in just 79 steps using the force relaxation approach of Sellers \cite{sellers}.
In these algorithms, strong damping \cite{sellers} or line searching \cite{sclegel_linesearch}
is required to avoid divergence. Another important technique to accelerate convergence of 
geometry optimization is the geometric DIIS (GDIIS) \cite{Pulay_GDIIS}. While geometric DIIS is a simple 
technique, its successful application may require many intricate modifications \cite{Farkas_GDIIS}.  A 
recent review and comparison of these algorithms and their technical details can be found in 
Ref.~\onlinecite{bakken}.  

\section{A New Concept} \label{concept}

In the Newton-Raphson (NR) approach to geometry optimization, a correction $\delta {\bf x}$  to the 
Cartesian vector $\bf x$ of nuclear positions is found by solving a local second order expansion of the 
energy function, yielding  $\delta {\bf x}  = - {\bf g}\cdot {\bf h}^{-1}$,  where $\bf h$ 
is the Hessian, a matrix of energy second derivatives with respect to nuclear displacement,
and $\bf g$ is the gradient vector of energy first derivatives.  Thus, 
a step along a specific element of $\bf x$ is coupled to all components of $\bf g$ through off-diagonal 
elements of $\bf h$.  Alternatively, it is possible to work in the principal axis system of $\bf h$, or 
normal modes,  which derive from eigenvectors of the Hessian.  On an ideal harmonic surface the NR method 
remains exactly decoupled to all orders in the basis of normal modes, and also achieves quadratic convergence.  

Now consider an approximate NR algorithm in a basis of diagonalizing normal modes,  applied to a 
non-ideal anharmonic surface.  For the moment,  assume that the normal modes remain diagonalizing with each
geometry step, but the eigenvalues change.  Then, if the geometry steps are small,
eigenvalues can be re-computed with backward differences of the gradient information at each step. 
With infinitesimal steps, this independent coordinate approximation is formally equivalent to NR.  

Realistically though, both eigenvectors and eigenvalues of the Hessian will change as the 
algorithm takes the largest possible steps, leading to second (and higher) order coupling effects.  At this point, 
we turn to an internal coordinate system to continue development of the independent coordinate 
approximation.  In an internal coordinate system, the magnitude of harmonic and anharmonic coupling
effects are small over the course of optimization.  On the other hand,
in a static basis of normal modes anharmonic coupling effects are larger to begin with \cite{fogarasi_diaghess}, 
and coupling will increase during the course of optimization unless the normal modes are 
periodically recomputed, incurring an unacceptable cost for large systems.  

Extending the independent coordinate approximation, it is clear that a differencing scheme will be inaccurate 
due to large step sizes.  Also, because coordinate independence is an approximation, coupling effects will lead
to scatter in the gradient data.  Replacing  the differencing scheme with a curve fit has the potential to both 
smooth the data and provide higher order accuracy.  Using the local tangent of a fitted curve to approximate 
diagonal elements of the Hessian 
would be similar to force relaxation \cite{pulay_review,sellers,van_alsenoy_98}, but with constantly updated 
values of the Hessian.  However, the curve fit can be carried further;  forsaking an approximate Newton step, 
the curve fit can be used instead to extrapolate or interpolate a zero for the parameterized gradient data.  
To first order (a line fit), this is equivalent to a one-dimensional Newton step with a fitted gradient and a 
fitted Hessian.   While the approach outlined so far works well, the use of standard least squares is not 
optimal due to the ability of outliers to skew the fit.  A significant advance can be realized by instead using 
a weighted curve fit, with the weights depending indirectly on estimation of local coupling effects. 

We call this concept the ``QUasi Independent Curvilinear Coordinate Approximation'' or QUICCA. 

Explicitly then, application of QUICCA to geometry optimization involves the  projection of Cartesian 
gradient information onto an uncoupled $N_{\rm int}$ dimensional internal coordinate space.  Energy gradients 
are accumulated for each internal coordinate, as shown in Fig.~\ref{NH3outp6}, and used to formulate 
$N_{\rm int}$ independent curves by weighted least squares.  At each step, these curves are extrapolated 
(or interpolated) to zero, with collective changes in the internal coordinate system taken as an approximate 
minima for the system as a whole. The final back-transformation of the step to Cartesian coordinates merges 
these displacements, reconciling any redundancies in the internal coordinate system. The quality of this 
collective extrapolation is determined by severity of the low order coupling, redundancy of the internal 
coordinate system and the accuracy of the fit.  The QUICCA approach to geometry optimization is simple to 
implement and scales linearly with system size.  It overcomes low order coupling effects through averaging and 
weighting of the curve fit and also admits the accumulation of anharmonic information.  However,  augmenting
QUICCA with anharmonic quadratic and higher order terms {\em and} preserving the important smoothing effect of 
the weighted fit is an advanced subject beyond the scope of this work. 

\section{Implementation}\label{implementation}

\subsection{Recognition of Internal Coordinates} \label{recognition}

The present implementation of QUICCA employs a redundant set of primitive internal 
coordinates. Similar coordinate systems are used by a majority of other internal coordinate optimizers 
\cite{Pulay_natural_internals,fogarasi_diaghess,bakerstest,eckert,schlegel_on2iter}.  
The generation of these coordinates starts with the recognition of covalent bonds, 
based on atomic Slater radii \cite{SlaterRad}  multiplied by an empirical factor of 1.3. 
Weak bonds are recognized on the basis of atomic van der Waals (VDW) radii \cite{JMOLtable}.
Note, that no VDW contact is allowed between atoms, which are second or third neighbors in the 
covalent bonding scheme.  The search for neighboring atoms is carried out by dividing 
the system into small cubes a few angstroms on a side, and bonds are looked for only within 
each box and between nearest neighbor boxes.  This divide-and-conquer scheme ensures the fast 
and linear scaling recognition of bonding for large molecules.   The bonding scheme is then 
stored as the symbolic component (graph) of the sparse topology matrix.

At each step of the optimization, the  bonding scheme is determined and the corresponding 
primitive topology matrix is stored.  Then, primitive topology matrices from the 
$n_{\rm fit}$ most  recent steps are summed to yield a merged topology matrix, where 
$n_{\rm fit}$ is the same number of points employed by the curve fitting.   
In the current implementation $n_{\rm fit}=7$, but it can be much larger or smaller.  

The merged topology matrix avoids fluctuations due to the possibility of bond-formation and 
bond vanishing during the optimization of large molecules.  Use of the merged topology matrix
increases the number of internal coordinates and the redundancy of the coordinate set only slightly;
the final merged topology matrix is determined directly from primary topology matrices rather 
than the merged ones at previous cycles.

Based on the topology matrix of the merged bonding network, all other primitive 
internal coordinates, such as bond angles, torsions, linear bendings and 
out-of-planes are recognized. Torsions are selected so that each bond will be 
associated with only a single torsional coordinate.

\subsection{Preparing the Data}

Cartesian coordinates and the corresponding Cartesian gradients of the $n_{\rm fit}=7$ 
most recent geometries are read from disk. Based on the most recent definition of the 
bonding scheme they are transformed into the internal coordinate representation,
determined by the merged bonding scheme.  The coordinate transformation is carried out as 
described in Ref.~\onlinecite{nemeth_coordtrf1}, while the necessary intermediate sparse matrices of 
the coordinate transformation are re-generated. This is a very inexpensive step, due to the 
extreme sparsity of the vibrational B matrix \cite{wilson}. 
Thus the primary objects of the proposed geometry optimization,  the internal coordinates and their gradients,
are re-created at each step allowing on-the-fly redefinition of the 
coordinate system.

In order to avoid problems due to the periodicity of torsional angles, reference points are used, 
eg.~the torsional angle values of the most recent geometry. Each torsional angle is set to its 
periodic equivalent closest to the reference point, by a shift of $\pm 2\pi$, while its gradient value 
does not change.

\subsection{The Weighted Curve Fit} \label{Sweights}

At this point,  we have a set of $n_{\rm fit}$ coordinate - gradient pairs for each individual 
internal coordinate, such as the one presented in Fig.~\ref{NH3outp6}. These data 
pairs parametrize progression of the gradient during the $n_{\rm fit}$ most recent steps of the 
optimization.  This data is now used to fit curves for each of the $N_{\rm int}$ coordinates using the 
{\sc SLATEC} routine {\sc POLFIT} and its dependencies \cite{slatec}, which perform a weighted least squares fit.

The choice of weight is a key factor in the behavior of the QUICCA algorithm, as it is used to prioritize 
reliable portions of the data and to generally smooth noisy trends.  
Originally, we considered the standard expression $w_{k}^{(i)}=\left(1/{g_{k}^{(i)}}\right)^2$ for the weight 
of the $k$-th internal coordinate at the $i$-th geometry, which is generally recommended for use in curve 
fitting \cite{numerical_recipies}.  These weights however do not 
account for the fact that the gradient of the $k$-th internal coordinate may be small, while 
its immediate  environment may contain large gradients indicating the possibility of large changes in the $k$-th 
internal coordinate due to coupling effects.  Thus, a data point with high environmental tension 
is less reliable than a point with neighboring coordinates that are relaxed.   
Local sums of the gradient squares,
\begin{equation}
\label{weights}
w_{k}^{(i)} = \left[ \sum_{l} g_{l}^{(i)} \frac{1}{|H_{ll}^{}|} g_{l}^{(i)} \right]^{-1} ,
\end{equation}
incorporate this effect, where  $l$ indexes internal coordinates that have atoms in common with 
the $k$-th internal coordinate, and $H_{ll}^{}$ is an approximate, static approximation to the 
diagonal Hessian.  The terms $H_{ll}$ yield dimensionally consistent sums, since each of the gradient 
components  $g_{l}$ has a different measure and energy scale for stretches, bendings and torsions.  

For each coordinate $k$, the sum over $l$ is determined by the topology matrix, given by the symbolic 
structure (graph) of the sparse matrix $G_{i}=BB^{t}$, which obtains from the sparse Wilson's $B$ 
matrix \cite{wilson}.  We have considered more delocalized interactions as well, such as those deriving 
from the graph of $G_i^2$, but they did not provide any significant improvement in test calculations.

Values of $1.0$ au for stretches, $0.1$ au for bendings (and linear bendings and out-of-planes) and $0.01$ au
for torsions are used for the $H_{ll}$.  These values are similar to those used to provide 
an initial guess for the Hessian in variable metric algorithms \cite{bakken}.  Note, that only the relative 
values of $H_{ll}$ are important for the fitting process, as the fitting parameters are determined by the 
relative values of the weights.  

While it is possible to employ second and third order polynomials in the fit, in our 
current implementation extrapolation to zero is most robust with just a line.  
A primary difficulty in the construction of higher order fits is that distinguishing 
between noise and signal is more difficult.  Also, there are stability 
issues of extrapolation versus interpolation with higher order polynomials.
Thus while we believe there is opportunity for higher order approximations,  
their development is beyond the scope of the present paper.  

\subsection{Minimum or Transition State?}

Transition states are indicated by a gradient curve with  negative tangent at the position
of the root.  For example, in Fig. \ref{NH3outp6} the tangent of the gradient curve is negative 
in the vicinity of the planar structure (structure \#1), which is a transition state. The tangent 
of the gradient curve thus offers a convenient tool to control convergence to either a minimum 
or transition state.  Whenever the local tangent of the gradient curve is negative, the optimizer 
employs a simple force relaxation step,
\begin{equation}
\label{tseq}
\varphi_{k}^{(i+1)} = \varphi_{k}^{(i)} -g_{k}^{(i)}/|H_{kk}^{(i)}| ,
\end{equation}
where $i$ denotes the serial number of the optimization step,  $\varphi_{k}$ is the $k$-th internal coordinate,
 $H^{(i)}_{kk}$ is a diagonal element of the Hessian determined from local tangent of the curve fit, and
$g_i$ is the local gradient at $i$ (not a fitted gradient).
This avoids transition states that are well localized to just a few internal coordinates.
The ability of the present implementation to treat delocalized transition states
remains to be seen. 

Combined internal coordinates, such as the natural \cite{Pulay_natural_internals} or the
delocalized internals \cite{Baker_deloc_1}, hold the possibility of identifying delocalized transition 
states by gradient fitting.  An analysis by  Wales {\it et.al.} \cite{Wales_saddlepoint} provides 
insight into the numerical problems involved in controlling convergence to a minimum without an 
accurate Hessian.  Note that, even with a good quality Hessian update, such as BFGS \cite{RFletcher},
there is no guarantee to converge into a minimum instead of a transition state. 
This is because positive definiteness of an approximate Hessian does not guarantee
the positive definiteness of the exact Hessian.  Thus, the exact Hessian remains the only 
absolute for determining the character of extremal points in the potential energy 
surface \cite{Pulay_natural_internals}.

\subsection{Initial steps}

In the very first step of the optimization, a force relaxation step is used with the usual 
diagonal Hessian estimates of 0.5 au for stretchings, 0.2 au for bendings and 0.1 au for 
torsions.  This initial step is equivalent with the one traditional optimizers use.

\subsection{Step size control and backtracking}

Experience with QUICCA has shown that extrapolations larger 
than the range spanned by the fitted data points are unreliable.   Thus, 
while QUICCA  always tries to achieve the predicted minimum, 
stepsize control is imposed, limiting the step  by  the fit range or to 
0.3 au, whichever is less.   This maximum allowed value of the stepsize, 0.3 au, 
is similar to that used by other optimizers \cite{eckert}.  

The range spanned by the fitted data is in some sense an uncertainty.  
Refining and formalizing quantification of this uncertainty is another area of 
opportunity for improvement of the QUICCA optimizer, which is beyond the 
scope of this article.  We note however, that it might be more effective
to use a filtered range of the data points, based on their weights. 
For example, compute the range including only those points with weights that 
comprise, say,  99.99 percent of the total.   Alternatively, statistical 
tests could be employed to filter out outlying points.  We note that issues of
uncertainty in stepsize control are certain to be far more complex for higher 
order fits, and advanced statistical methods may have a significant role 
to play in this area of development.

\subsection{Backtracking}

While the above stepsize control is designed to avoid potential instabilities,
it is still possible to take steps that will increase the total energy.  In 
this case, there are a number of options.  We can live with the increase, 
hoping it will nevertheless improve the information content used to build the energy 
surface.  We can backtrack, uniformly reducing the step, typically by half, and re-evaluate
the energy to see if it decreases.   Alternatively, we can employ a more 
sophisticated and expensive line search to determine a minimizing step.  

There are many ways to carry out a line-search.  Perhaps the most widely employed 
approach is the endpoint algorithm of Schlegel \cite{sclegel_linesearch}.  This algorithm
employs energies and gradients at the endpoints to yield a quartic interpolant.  This
approach will work well if the interpolant varies smoothly between endpoints.  We have
implemented this approach to the line-search, but found that it does not perform 
reliably, especially when large steps are made.   It may be that midpoint algorithms
are more appropriate, but we have not explored this further.

Our experience with QUICCA is that simple back-tracking is both cheaper and more effective
than the endpoint line-search.  During back-tracking, a  bad step is recognized by an elevated
energy before any gradient evaluation takes place. Then, the step-size is halved,
and a new energy is calculated.  This process is repeated until the energy becomes lower,
typically requiring just one halving.

\subsection{Iterative back-transformation}\label{transformation}

Once predictions of internal coordinate displacements are made, they are passed to the 
iterative back-transformation to generate a new set of Cartesian coordinates, as in 
Ref.~\onlinecite{pulay_review}.   However, there is no guarantee that there exists a set of 
Cartesian coordinates that can exactly realize the new, predicted internal coordinates.  
For example, a prediction of angles in a triangular molecule can sum 
to a value greater than or less than 180 degrees. Such conflicts are a consequence of 
the redundant internal coordinate system.  

Displacements in the updated internal coordinate system that result in geometric conflicts 
will be projected out during the iterative back-transformation. However, if 
these conflicts are too large, the iterative back-transformation may fail to converge,
or it may produce a set of Cartesian coordinates that violates the predicted internal 
coordinate values. Typically, these violations are big when stepsizes are large, but become 
small close to convergence.  

One approach to avoiding this problem is to work with linear-combinations of primitive internal
coordinates, such as the natural \cite{Pulay_natural_internals} or delocalized
\cite{Baker_deloc_1} internal coordinates, which minimize geometric conflicts.  
Another cure for redundancies is to apply a projection technique prior to the iterative-back 
transformation \cite{pulay_review}, which can eliminate problems to first-order, but does not 
guarantee convergence for large steps.

In our present implementation, geometric conflicts are dealt with solely by the iterative back-transformation.
In the case of non-convergent back-transformation,  our algorithm will attempt the back-transformation
again, but with the original internal coordinate displacements halved.  If the repeated transformation 
also fails, it usually indicates an incomplete set of internal coordinates and the program stops.  An incomplete 
set can also scuttle convergence of the Cartesian to internal gradient transformation \cite{nemeth_coordtrf1}. 
Note, that an appropriate internal coordinate recognition program 
together with suitable internal coordinate step-size control
can always avoid the non-convergent back-transformation.

\subsection{Realization}

The QUICCA geometry optimization algorithm has been implemented in the
object oriented FORTRAN-90/95 programming language and is part of
the MondoSCF suite of linear scaling quantum chemistry codes \cite{MondoSCF}.
It has been successfully tested on different platforms, compilers and operating
systems. 

\subsection{Scaling}

At each geometry step, the cost of QUICCA is associated only with 
recognition of internal coordinates, the coordinate 
transformations and the fitting process. The recognition of internal 
coordinates scales linearly, as described in Section IV A.  
The effort spent on fitting is $ {\cal O}(N_{\rm int})$ with $N_{\rm int} \propto N_{\rm atoms}$, as the number of 
points employed by each fit is a constant.  Likewise, the coordinate transformations scale linearly with system 
size as pointed out in Ref.~\onlinecite{nemeth_coordtrf1}.

\section{Results and Discussion} \label{Sresults}

\commentoutA{

\begin{table}[h]
\caption{
Number of geometry optimization steps to convergence of Baker's test set
(up to 30 atom molecules) for the QUICCA optimizer and also for other 
advanced algorithms from the literature.  Calculations have 
been carried out at the RHF/STO-3G level of theory.  All of the reported
calculations used local primitive internal coordinates, except those
of Bakken {\it et.al.} who used extra redundant coordinates \cite{bakken}.
}
\label{Bakers_test}
\begin{tabular}{lccccc}
\toprule
Molecule               & QUICCA  & Bakken         & Eckert         & Lindh         &  Baker  \\
                       &         & {\it{et al.}}  & {\it{et al.}}  & {\it{et al.}} &    \\
                       &         &  \cite{bakken} &  \cite{eckert} & \cite{lindh}  &  \cite{bakerstest} \\
\colrule
Water                  &   4    &   4    &    4    &    4   &   6     \\
Ammonia                &   4    &   5    &    6    &    5   &   6     \\
Ethane                 &   4    &   3    &    4    &    4   &   5     \\
Acetylene              &   4    &   4    &    6    &    5   &   6     \\
Allene                 &   3    &   4    &    4    &    5   &   5     \\
Hydroxysulphane        &   8    &   7    &    7    &    8   &   8     \\
Benzene                &   2    &   3    &    3    &    3   &   4     \\
Methylamine            &   4    &   4    &    5    &    5   &   6     \\
Ethanol                &   5    &   4    &    5    &    5   &   6     \\
Acetone                &   4    &   4    &    5    &    5   &   6     \\
Disilyl ether          &   9    &   8    &    9    &   11   &   8     \\
1,3,5-trisilacycl.     &   5    &   9    &    6    &    8   &   8     \\
Benzaldehyde           &   5    &   4    &    5    &    5   &   6     \\
1,3-difluorobenz.      &   4    &   4    &    5    &    5   &   5     \\
1,3,5-trifluorob.      &   3    &   4    &    4    &    4   &   5     \\
Neopentane             &   5    &   4    &    4    &    5   &   5     \\
Furan                  &   6    &   5    &    6    &    7   &   8     \\
Naphtalene             &   6    &   5    &    6    &    6   &   5     \\
1,5-difluoronapht.     &   6    &   5    &    6    &    6   &   6     \\
2-hydroxibicyclop.     &   7    &   9    &    9    &   10   &  15     \\
ACHTAR10               &  10    &   8    &    9    &    8   &  12     \\
ACANIL01               &  11    &   7    &    8    &    8   &   8     \\
Benzidine              &   8    &   9    &    7    &   10   &   9     \\
Pterin                 &  12    &   8    &    9    &    9   &  10     \\
Difuropirazine         &   6    &   6    &    7    &    7   &   9     \\
Mesityl oxide          &   5    &   5    &    6    &    6   &   7     \\
Histidine              &  11    &  16    &   14    &   20   &  19     \\
Dimethylpenthane       &   6    &   9    &   10    &   10   &  12     \\
Caffeine               &   7    &   6    &    7    &    7   &  12     \\
Menthone               &  13    &  12    &   10    &   14   &  13     \\
\colrule
Total                  & 187    & 185    &  196    &  215   & 240     \\
\botrule
\end{tabular}
\end{table}

}

\subsection{Baker's test set}

The performance of the QUICCA optimizer has been tested on Baker's suite \cite{bakerstest} of 
small test molecules, and is listed in Table~\ref{Bakers_test}, along with values from more traditional 
algorithms.  At convergence, the magnitude of the maximum Cartesian force vector is required to be less 
than $3\times10^{-4}$ au for each atom.  This criterion is tighter than that employed in other studies 
Ref. \onlinecite{bakerstest}, which require only that the maximum Cartesian force component (X, Y or Z) be less 
than $3\times10^{-4}$ au.  In addition to this criterion,  an absolute energy change less than 
$1\times10^{-6}$ au or a predicted internal coordinate displacement less than $3\times10^{-4}$ au at the last 
geometry is required.  The energies of the optimized structures have been reproduced to all digits given 
in Ref.~\onlinecite{bakerstest}. 

The current implementation of QUICCA, achieving a total of 187 geometry steps for Baker's test suite, is 
competitive with Bakken's algorithm, requiring 185.  

Bakken's method \cite{bakken} derives from an optimal combination of advanced techniques
related to internal coordinate methods based on a dense Cartesian Hessian update. It involves 
a Cartesian BFGS update \cite{RFletcher}, the Rational Function Optimization scheme \cite{benerji_RFO} 
and a Newton trust-region model \cite{RFletcher}. An important addition to these techniques is the 
extra redundant coordinates, which contribute to the success of Bakken's algorithm relative to more 
traditional schemes. Extra redundant internal coordinates are local distance coordinates, which bridge 
the terminal atoms of all valence angles and most of the dihedral angles.

In contrast, the current implementation of QUICCA employs a redundant primitive internal coordinate 
system, a weighted line fit and simple backtracking.  

In two cases (ACANIL01 and Pterin), the QUICCA optimizer proved to be considerably slower 
than the best traditional techniques.   We believe these cases may be improved by introducing a
more sensitive weighting scheme, and/or employing a more sophisticated internal coordinate
system as discussed above in Section IV H.

Note that these results have been achieved without the use of traditional convergence acceleration 
algorithms such as geometric DIIS (GDIIS) \cite{Pulay_GDIIS,Farkas_GDIIS}.  It is of course entirely 
possible to deploy the GDIIS algorithm together with QUICCA, but we leave this avenue open for future 
work.

Our experience shows, that QUICCA is relatively insensitive to the values of $H_{ll}$ employed 
in the construction of weights (see Section IV C).   For example the values $0.5$, $0.2$ 
and $0.1$ au for $H_{ll}$ (instead of our standard, $1.0$, $0.1$ and $0.01$ au)
resulted in 189 steps instead of 187 for Baker's test set.  The largest difference was present 
in Dimethylpenthane, which converges 2 steps faster when using the standard $H_{ll}$ values. 

In  Baker's test suite, about three force evaluations have been saved by 
using backtracking. Without the application of backtracking, Baker's test set
converges in 190 steps.  This suggests that the core QUICCA algorithm is robust and
does not often overstep.  

We can understand the stability achieved by the core QUICCA
algorithm by considering a hybrid force relaxation method, based on Eq.~\ref{tseq},
in which the diagonal Hessian elements $H^{(i)}_{kk}$ are obtained from the 
curve fit, but the uncoupled Newton step is carried out with the {\em current} 
gradient.  We have tested this alternative, but it turned out to be oscillatory 
in many cases, requiring excessive backtracking.  This should be contrasted with the
full QUICCA algorithm, which uses extrapolation. To first order this is equivalent with 
an approximate Newton step taken with the {\em fitted} gradient.   Thus, the full 
QUICCA algorithm is successful not only because it generates up-to-date curvature information, 
but because it achieves stability through smoothing effects derived from the weighted fit.

\commentoutA{

\begin{figure}[h]
\resizebox*{3.5in}{!}{\includegraphics{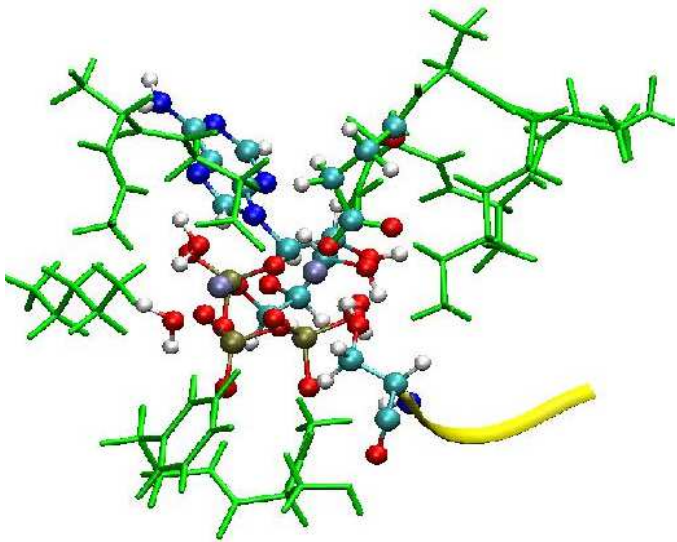}}
\caption{
The structure of a 263 atom Protein Kinase A fragment used
to test the QUICCA optimizer.}\label{kinasepicture} 
\end{figure}

\begin{figure}[h]
\resizebox*{3.5in}{!}{\includegraphics{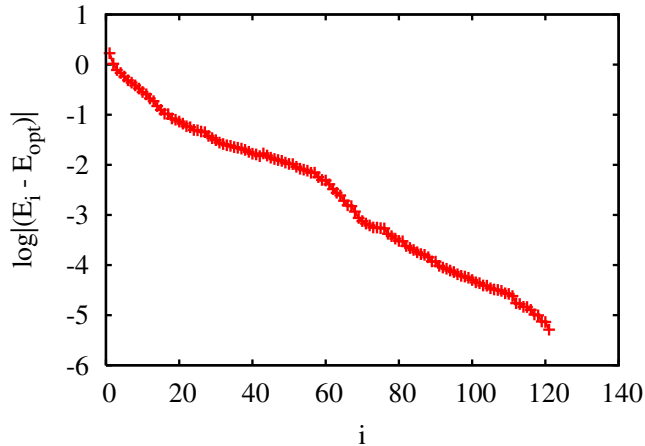}}
\caption{
Convergence of the energy with geometry step $i$ on a log$_{10}$-linear  scale for the 
263 atom Protein Kinase A fragment at the RHF/STO-2G level of theory.}\label{logn-logde} 
\end{figure}

\begin{figure}[h]
\resizebox*{3.5in}{!}{\includegraphics{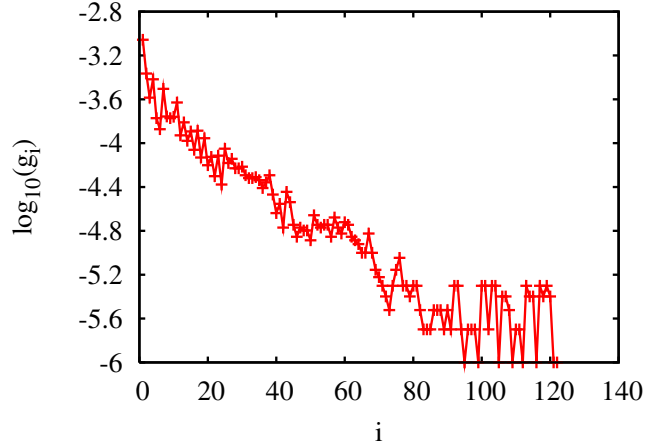}}
\caption{
Convergence of the root mean square of the Cartesian gradients, $g_{i}$,
with geometry step $i$ on a log$_{10}$-linear  scale for the 
263 atom Protein Kinase A fragment at the RHF/STO-2G level of theory.}\label{gradientpicture} 
\end{figure}

\begin{figure}[h]
\resizebox*{3.5in}{!}{\includegraphics{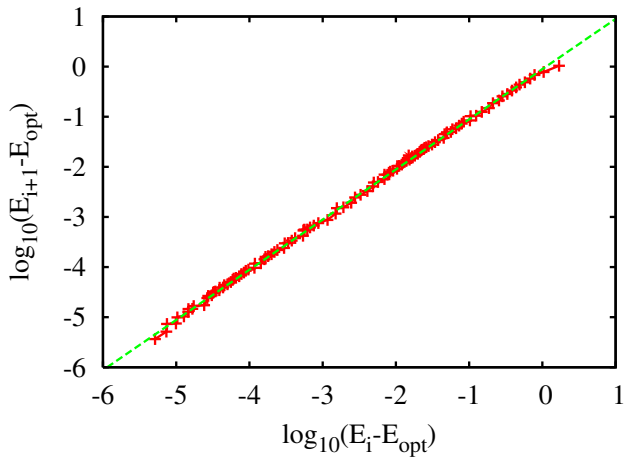}}
\caption{A plot of ${\rm log}_{10} \Delta E_{i+1}$ vs  ${\rm log}_{10} \Delta E_{i}$,  characterizing the ratio 
         $\Delta E_{i+1}/\Delta E_i$.  The solid line is $y=-0.04980+1.00111 *x$. 
         The slope  of this line is equivalent to the order of convergence ($n=1.00111$), 
         while the y-intercept is the negative rate of convergence ($-f ={\rm log}_{10}(c)= -0.04980$).}
\label{loglogplot}
\end{figure}

\begin{figure}[h]
\resizebox*{3.5in}{!}{\includegraphics{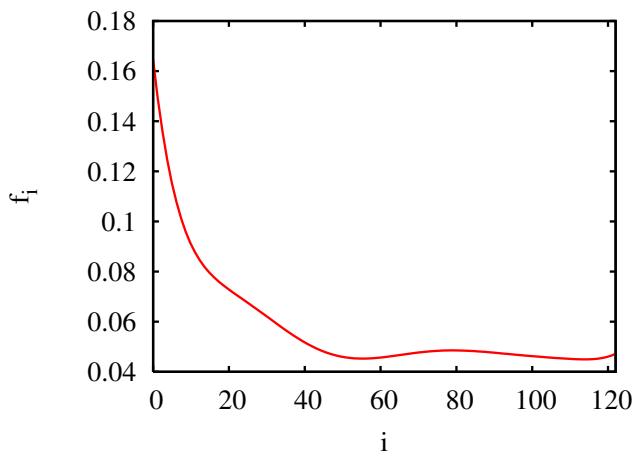}}
\caption{
Change in the rate of convergence $f$ during  optimization
of the 263 atoms Protein Kinase A fragment. $f$ is modeled
by a 10th order polynomial. The polynomial coefficients
have been obtained from a non-linear fit to the $E_{i}$-$i$
data-pairs, using the model function $E_{i}=E_{\rm opt}+A*e^{-f*i}$.}
\label{convfact} 
\end{figure}
}

\subsection{Convergence}

Away from the basin of attraction, both steepest descents as well as the 
Newton-Raphson algorithm converge linearly.   However, these methods behave very differently
when condition number of the Hessian is large, for example when low frequency ``floppy'' modes
are present.  In this case, steepest descents can be extremely slow, while second order 
(Newton-Raphson) and approximate second order (quasi-Newton) methods achieve an accelerated rate 
of convergence through conditioning of the stepsize.   While methods that render  the
number of geometry steps independent of system size do exist \cite{SGoedecker01}, they apply (so far) 
only to homogeneous systems.  In general though, the number of steps required to minimize to within
a certain tolerance increases with system size, roughly as ${\cal O}(N_{\rm atoms}/3)$ 
according to Ref.~\onlinecite{Schlegel_plasminogen_opt}.  Therefore, the rate of linear convergence achieved by an 
optimization algorithm is perhaps a good  measure of performance for large systems.  

An optimization algorithm may be characterized by the ratio ${\Delta E_{i+1}}/{\Delta E_{i}^n} =c$,
where $\Delta E_i = |E_{i}-E_{\rm opt}| $ is the absolute error at geometry step $i$,
$E_{\rm opt}$ is the extremal value of the energy, $n$ is the order of convergence, 
and $c$ is the convergence factor \cite{Quarteroni}.  With first order convergence $n=1$, and
the error typically decays exponentially as $\Delta E_i = e^{-f_i*i}$, where $f=-{\rm log}(c)$ is the 
rate of convergence. In the basin of attraction, where the second order approximation 
to the nonlinear objective is accurate, the Newton-Raphson method achieves second order convergence,
$ n=2$, while approximate second order methods (quasi-Newton algorithms for example) can 
at best achieve superlinear convergence \cite{RFletcher,Pulay_natural_internals}.   Superlinear 
convergence is characterized by $n=1$ and an increasing rate of convergence $f$, equivalent to 
$\lim_{i \to \infty} c_i = 0$.  

To investigate the convergence properties of QUICCA in a large application, we have carried out
optimization of a 263 atom fragment of crystallographic Protein Kinase A, involving two Magnesiums,
three crystallographic waters, ATP and a fragment of the Protein Kinase Inhibitor to which phosphate 
is transfered. This is a model reactant system obtained from Ref.~\onlinecite{graeme} and is shown in 
Fig.~\ref{kinasepicture}.  Optimization was performed at the HF/STO-2G level of theory with an 
approximate absolute energy resolution of $10^{-5}$ hartree using MondoSCF \cite{MondoSCF} at the 
{\tt tight} accuracy level and with ten peripheral atoms  constrained to mimic steric effects of the enzyme.  
Note that this minimal level of theory actually makes more work for the optimizer, as it will tend to 
move away from the more correct crystallographic structure.   This structure was optimized to within a 
Cartesian force of $3\times10^{-4}$ au and a maximum predicted displacement less than $4\times10^{-3}$ au.
The number of internal coordinates, generated for this structure was 1280 at convergence.  Figure~\ref{logn-logde} 
shows ${\rm log}_{10} \Delta E_i$ vs $i$ for this system, Fig.~\ref{gradientpicture} the root mean square 
of the Cartesian gradient at each step.

In Fig.~\ref{loglogplot}, a plot of ${\rm log}_{10}{\Delta E_{i+1}}$ vs ${\rm log }_{10} {\Delta E_{i}}$ 
is shown, demonstrating only first order convergence with a rate $f \sim 0.05$.   To examine the rate of convergence
in greater detail, we fit the the model function $E_{i}=E_{\rm opt}+A*e^{-i * f_i}$, using  
an 11th order polynomial to represent $f_i$.  The fitted rate of convergence shown in Fig.~\ref{convfact}
suggests that the optimization has three different phases, in accordance with Fig.~\ref{logn-logde}.
In the first phase (steps 1-10) there is a rapid refinement of the least-optimal substructures. In the 
second phase (steps 10-60) the absolute energy decreases less rapidly to $\sim 1$ mhartree, with the  
main optimization effort corresponding to changes in the local internal coordinates. Finally, the third phase 
(60-120) occurs at a slower, constant rate of about $f=0.05$, which is consistent with the fit characterizing 
the ratio ${\Delta E_{i+1}}/{\Delta E_{i}}$ shown in Fig.~\ref{loglogplot}.

In the third phase, below 1 mhartree, backtracking becomes necessary to achieve smooth 
convergence of the energy and the rate of convergence decreases to a constant.  We believe
that this deceleration is due to the presence of ``floppy'' low-frequency modes
that are highly delocalized and not accounted for in the current local internal 
coordinate system.  While QUICCA can certainly be used with a more delocalized 
internal coordinate system, the construction of delocalized internal coordinates 
is an important open problem in molecular geometry optimization.   It is also important to note that 
the effect of floppy modes is not the only factor in the third phase of the optimization; the rigidity 
of the internal coordinate system and its role in the inaccurate convergence of the back-transformation 
is also important. In our opinion, these two factors, which characterize the last phase of the 
optimization may be related to each other.  A recent work of Fernandez-Serra {\it et.al} \cite{Serra03} 
calls attention to the relationship of the rigidity of internal coordinate systems and the 
phenomenon of floppy modes. This relationship may be understood from the theory of 
rigidity \cite{Phillips85}. In the present work, we do not 
consider this relationship further.

\section{Conclusions}

We have presented a new concept for local optimization, the ``QUasi Independent Curvilinear 
Coordinate Approximation'' or  QUICCA, and applied it to the geometry optimization of molecular 
structures. Competitive with the most aggressive internal coordinate geometry optimizers in the literature, 
QUICCA has demonstrated that an efficient optimization algorithm can be constructed without 
{\em explicit} coupling in local expansion of the energy.  However, coupling effects do enter 
QUICCA {\em implicitly} through the weighted curve fit, providing an important averaging effect that 
compensates for irregularities related to the independent coordinate approximation.  

The implementation presented here is relatively simple, employing redundant primitive 
internal coordinates, a weighted line fit and backtracking, allowing  substantial room for 
improvement.  Indeed, the QUICCA concept opens the door for a wide range of auxiliary techniques 
that may benefit from physical insight (such as the development of more precise weighting schemes) 
as well as from ideas in applied mathematics involving robust estimation, interpolation and 
advanced methods of regression.  

The primitive internal coordinate system employed in the present implementation does not include
delocalized motions and suffers from slowdowns associated with floppy modes.  The construction of 
delocalized internal coordinate systems that remain approximately
decoupled is thus an outstanding problem in molecular geometry optimization.   Nevertheless, the
preconditioning afforded by local stretchings, bendings and torsions is an important step, enabling 
the large scale application of  linear scaling electronic structure theory to problems in biology and 
materials science.   In addition, with the availability of fast linear scaling Cartesian/internal coordinate 
transforms \cite{nemeth_coordtrf1}, the QUICCA algorithm may also present advantages in the optimization of 
large systems employing classical forcefields.  

The averaging scheme presented here, achieving an effective decoupled internal coordinate representation
of the potential energy surface, may also be useful in on-the-fly construction of classical force fields.
In such an approach, force field parameters such as the spring-constant and equilibrium position 
could be updated periodically from {\em ab initio} data, taken over several time steps. 
This on-the-fly construction of force fields based on QUICCA might provide a useful  alternative to 
the force matching algorithm of Ercolessi {\it et.al.} \cite{force-matching} and to the reactive 
force field concepts developed by the Goddard group \cite{reaxff1,reaxff2}.

\begin{acknowledgments}
K.~N{\'e}meth gratefully acknowledges {\"{O}}.~Farkas (Budapest) for providing the 
Cartesian coordinates of the Baker's test set,  B.P.~Uberuaga (Los Alamos) for 
calling our attention to Ref.~\cite{force-matching} and  J. {\'A}ngy{\'a}n
(Nancy), S. Lucidi (Rome) and V. Weber (Los Alamos) for discussions.

This work has been supported by the US Department of Energy 
under contract W-7405-ENG-36 and the ASCI project.  
The Advanced Computing Laboratory of Los 
Alamos National Laboratory is acknowledged.
\end{acknowledgments}

\bibliography{QUICCA}

\end{document}